\documentclass{amsart}

\usepackage{amsmath,amssymb,graphicx}
\usepackage{lineno}
\usepackage{subfigure}
\usepackage{yhmath}
\usepackage[margin=1in]{geometry}
\usepackage{amsaddr}

\newcommand{\eps}{\varepsilon}

\newcommand{\om}{\omega}
\newcommand{\prt}{\partial}

\newcommand{\R}{\mathbb{R}}

\newcommand{\sgn}{\mathrm{sgn}\,}
\newcommand{\rmd}{\,\mathrm{d}}
\usepackage{setspace}
\newcommand{\abs}[1]{\left| #1 \right|}
\newcommand{\Order}[1]{\mathcal{O}(#1)}
\DeclareMathOperator{\sech}{sech}
\newcommand{\Acal}{\boldsymbol{\mathcal{A}}}
\newcommand{\Acalt}{\tilde{\boldsymbol{\mathcal{A}}}}
\newcommand{\Lcal}{\mathcal{L}}
\newcommand{\Pcal}{\boldsymbol{\mathcal{P}}}
\newcommand{\Qcal}{\boldsymbol{\mathcal{Q}}}
\newcommand{\IP}[2]{\langle #1 , #2 \rangle}
\newcommand{\phib}{\overline{\phi}}

\newcommand{\pt}{\tilde{\phi}}
\newcommand{\kt}{\tilde{k}}
\newcommand{\wt}{\tilde{\om}}
\newcommand{\at}{\tilde{a}}
\newcommand{\qt}{\tilde{\mathbf{q}}}
\newcommand{\rt}{\tilde{\mathbf{r}}}
\newcommand{\ct}{\tilde{c}}
\usepackage{lipsum}
\usepackage{amsfonts}
\usepackage{epstopdf}
\usepackage{algorithmic}
\ifpdf
  \DeclareGraphicsExtensions{.eps,.pdf,.png,.jpg}
\else
  \DeclareGraphicsExtensions{.eps}
\fi

\usepackage{amsopn}

\title{Modulations of viscous fluid conduit periodic waves}

\author{Michelle D. Maiden, Mark A. Hoefer}
\address{Department of Applied Mathematics, University of
    Colorado, Boulder, CO}
\email{michelle.maiden@colorado.edu, hoefer@colorado.edu}
\begin{document}

\maketitle

\begin{abstract}
 In this work,
  modulation of periodic interfacial waves on a conduit of viscous
  liquid is explored utilizing Whitham theory and Nonlinear
  Schr\"{o}dinger (NLS) theory.    
  Large amplitude periodic wave modulation theory does not require
  integrability of the underlying model equation, yet in practice,
  either integrable equations are studied or the full extent of
  Whitham (wave-averaging) theory is not developed. The governing conduit equation is
  nonlocal with nonlinear dispersion and is not integrable.  Via a
  scaling symmetry, periodic waves can be characterized by their
  wavenumber and amplitude.  In the weakly nonlinear regime, both the
  defocusing and focusing variants of the NLS equation are derived,
  depending on the wavenumber.  Dark and bright envelope solitons are
  found to persist in the conduit equation.  Due to non-convex
  dispersion, modulational instability for periodic waves above a
  critical wavenumber is predicted.  In the large amplitude regime,
  structural properties of the Whitham modulation equations are
  computed, including strict hyperbolicity, genuine nonlinearity, and
  linear degeneracy.  Bifurcating from the NLS critical wavenumber at
  zero amplitude is an amplitude-dependent elliptic region for the
  Whitham equations within which a maximally unstable periodic wave is
  identified.  These results have implications for dispersive shock
  waves, recently observed experimentally.
\end{abstract}

\section{Introduction}
\label{sec:introduction}
Nonlinear wave modulation is a major mathematical component of the
description of dispersive hydrodynamics.  Dispersive hydrodynamics
encompasses the study of fluid-like media where dissipative effects
are weak compared to dispersion \cite{el_dispersive_2016}.  Solitary
waves and dispersive shock waves (DSWs) are typical coherent
structures.  Model equations include the integrable Korteweg-de Vries
(KdV) and Nonlinear Schr\"{o}dinger (NLS) equations as well as
non-integrable counterparts that are important for applications to
superfluids, geophysical fluids, and laser light.  Modulation theory
assumes the existence of a multi-parameter family of nonlinear,
periodic traveling wave solutions whose parameters change slowly
relative to the wavelength and period of the periodic solution under
perturbation.  Such variation is described by modulation equations.
In the weakly nonlinear regime, the NLS equation is a universal model
for the slowly varying envelope, incorporating both cubic nonlinearity
and dispersion.  In the large amplitude regime, the Whitham equations
\cite{whitham_linear_1974} describe slow modulations of the wave's
mean, amplitude, and wavenumber.  At leading order, they are a
dispersionless system of quasi-linear equations.

\begin{figure}
    \centering
    \includegraphics{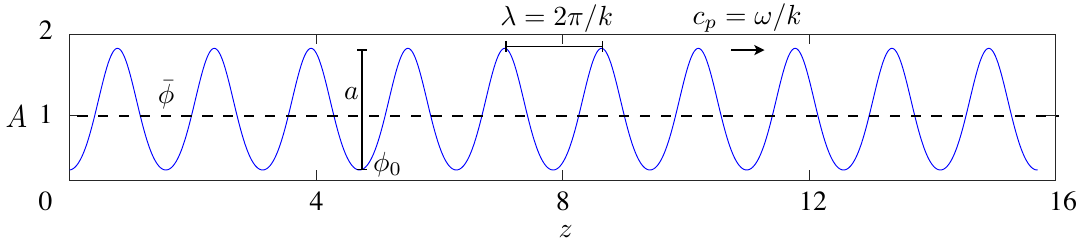}
    \caption{Example of a periodic wave solution to the conduit
      equation with wavenumber $k=2$, amplitude $a=1.5$, and unit mean.}
    \label{fig:periodic_wave}
\end{figure}

In this paper, we investigate nonlinear wave modulations in both the
weakly nonlinear and large amplitude regimes for the conduit equation
\begin{equation}\label{eq:conduit}
  A_t + (A^2)_z -(A^2(A^{-1}A_t)_z)_z = 0 .
\end{equation}
This equation approximately governs the evolution of the circular
interface, with cross-sectional area $A$, separating a light, viscous
fluid rising buoyantly through a heavy, more viscous, miscible fluid
at small Reynolds numbers
\cite{scott_observations_1986,lowman_dispersive_2013}.  Our motivation
for studying eq.~\eqref{eq:conduit} is two-fold.  First, the conduit
equation is not integrable \cite{harris_painleve_2006} so there are
mathematical challenges in analyzing its rich variety of nonlinear
wave features.  Second, equation \eqref{eq:conduit} is an accurate
model of viscous fluid conduit interfacial waves where hallmark
experiments have been performed on solitary waves
\cite{olson_solitary_1986,whitehead_magma_1990,helfrich_solitary_1990},
their interactions
\cite{scott_observations_1986,whitehead_wave_1988,lowman_interactions_2014},
and DSWs \cite{maiden_observation_2016}.  We believe the conduit
system is an ideal model for the study of a broad range of dispersive
hydrodynamic phenomena.  Indeed, in this work we elucidate additional
nonlinear wave phenomena predicted by eq.~\eqref{eq:conduit} by
analyzing the weakly nonlinear, NLS reduction and by examining the
structural properties of the large amplitude modulation (Whitham)
equations.  We use these equations to study properties of modulated
periodic wave solutions of the conduit equation, an example of which
is shown in Fig. \ref{fig:periodic_wave}

\subsection{Background on viscous fluid conduits}
\label{sec:backgr-visc-fluid}

Conduits generated by the low Reynolds number, buoyant dynamics of two
miscible fluids with differing densities and viscosities were first
studied in the context of geological and geophysical processes
\cite{whitehead_dynamics_1975}.  A system of equations describing the
dynamics of melted rock within a solid rock matrix was derived by
treating molten rock and its solid, porous surroundings as two fluids
with a large density and viscosity difference
\cite{mckenzie_generation_1984}.  Under appropriate assumptions, the
family of magma equations
\begin{equation}
  \label{eq:magma}
  \varphi_t + \left ( \varphi^n \right )_x - \left ( \varphi^n \left
      (\varphi^{-m} \varphi_t \right )_x \right )_x = 0
\end{equation}
describing the evolution of the volume fraction $\varphi$ of melted
rock, can be derived
\cite{richter_dynamical_1984,simpson_multiscale_2010}.  There are two
constitutive model parameters $(n,m)$ that relate the porosity of the
rock matrix to its permeability and viscosity, respectively.  The
conduit equation \eqref{eq:conduit} is precisely the magma equation
\eqref{eq:magma} when $(n,m)=(2,1)$ \cite{scott_observations_1986}.
Viscous fluid conduits, in contrast to magma, are easily accessible in
a laboratory setting, typically with a sugar solution or glycerine for
the exterior fluid, and a dyed, diluted version of one of these fluids
for the interior fluid
\cite{olson_solitary_1986,scott_observations_1986,whitehead_wave_1988,maiden_observation_2016}.

Early experiments primarily explored
the development of the conduit itself, which results in a diapir
followed by a periodic wavetrain
\cite{whitehead_dynamics_1975,olson_solitary_1986}. Solitary waves in
the established conduit have also been extensively studied, including
their amplitude-speed relation, interactions, and fluid transport
properties
\cite{olson_solitary_1986,scott_observations_1986,whitehead_magma_1990,helfrich_solitary_1990,lowman_interactions_2014}.
Experiments have also shown that interactions between solitons are
nearly elastic, with a phase shift the primary quantifiable change
\cite{helfrich_solitary_1990,lowman_interactions_2014}.  Furthermore,
soliton interaction geometries predicted by Lax for the KdV
equation \cite{lax_integrals_1968} were observed and agreed well with
numerical simulations \cite{lowman_interactions_2014}.  Recently,
dispersive shock waves (DSWs) were observed, yielding good agreement
with predictions from Whitham averaging theory
\cite{maiden_observation_2016}.  The accompanying observations of
soliton-DSW interaction suggest a high degree of coherence, i.e., the
sustenance of dissipationless/dispersive hydrodynamics over long
spatial and temporal time scales.  It is for this reason that we
further investigate modulations of periodic conduit waves.

\subsection{Properties of the conduit equation}
\label{sec:prop-cond-equat}

To fully describe the two-fluid interface of the conduit system, one
can consider the full Navier-Stokes equations with boundary conditions
along a moving, free interface.  However, in the small interfacial
slope, long-wave regime, a balance between the viscous stress force of
the exterior fluid and the buoyancy force acting on the interior fluid
leads to the asymptotically resolved conduit equation
\eqref{eq:conduit} with no amplitude assumption
\cite{olson_solitary_1986,lowman_dispersive_2013}.  The force balance
is achieved with small interfacial slopes on the order of the square
root of the ratio of the interior to exterior fluid viscosities.

The conduit equation \eqref{eq:conduit} has been studied since the
1980s.  The equation has exactly two conservation laws
\cite{barcilon_nonlinear_1986,harris_conservation_1996}:
\begin{equation}\label{eq:ConsLaws}
    \begin{cases}
    A_t + (A_2-A^2(A_{-1}A_t)_z)_z&= 0, \\[3mm]
    \displaystyle \left(\frac{1}{A} + \frac{A_z^2}{A^2}\right)_t +
    \left(\frac{A_{tz}}{A}-\frac{A_zA_t}{A^2}-2\ln{A}\right)_z &=0 .
    \end{cases}
\end{equation}
The conduit equation itself corresponds to conservation of mass and
obeys the scaling invariance
\begin{equation}\label{eq:scaling1}
  \begin{array}{ccc}
    \tilde{A}=A/A_0, & \tilde{z}=A_0^{-1/2}z,  & \tilde{t} = A_0^{1/2}t.
  \end{array}
\end{equation}
The linearization of the conduit equation upon a unit area background
admits trigonometric traveling wave solutions subject to the frequency
dispersion relation
\begin{equation}\label{eq:lineardispersion}
    \om_0(k) = \frac{2k}{1+k^2},
\end{equation}
with wavenumber $k$.  This leads to the linear phase $c_{\rm p}$ and
group $c_{\rm g}$ velocities
\begin{equation}\label{eq:linearvelocities}
  c_{\rm p}(k) = \frac{\om_0(k)}{k} = \frac{2}{1+k^2}, \quad  c_{\rm
    g}(k) = \om_0'(k) = \frac{2(1-k^2)}{1+k^2} .
\end{equation}
Note that $c_{\rm g} < c_{\rm p}$ for $k > 0$.  While the phase
velocity is always positive, the group velocity is negative for $k>1$.
Failure of the Painlev\'{e} test suggests that the conduit equation is
not completely integrable \cite{harris_painleve_2006}.  The conduit
equation is globally well-posed for initial data $A(\cdot,0) - 1 \in
H^1(\R)$ with $A(z,0)$ physically-relevant initial data bounded away
from zero in order to avoid the singularity
\cite{simpson_degenerate_2007}.

Solitary waves have been studied numerically for the more general
magma equation \eqref{eq:magma} where it has been found that solitary
waves exhibit near-elastic interactions resulting in a phase-shift and
a physically negligible dispersive tail
\cite{scott_magma_1984,nakayama_compressive_1991,lowman_interactions_2014}.
The asymptotic stability of solitary waves has also been proven
\cite{simpson_asymptotic_2008}. General (unmodulated) periodic wave
solutions have been found, and an implicit dispersion relation has
been computed for these waves \cite{olson_solitary_1986}.  In the long
wavelength, small amplitude regime, the conduit equation reduces to
KdV \cite{whitehead_korteweg-devries_1986}.

There have been several works applying periodic traveling wave
modulation theory to the magma equations \eqref{eq:magma}.  Reference
\cite{marchant_approximate_2012} considered Eq.~\eqref{eq:magma} with
$(n,m) = (3,0)$, describing DSWs and structural properties of the
Whitham equations.  Modulations of periodic traveling waves in the
magma equation \eqref{eq:magma} and a generalization of it were
investigated in the weakly nonlinear, KdV regime
\cite{elperin_nondissipative_1994}.  Modulated periodic waves in the
form of DSWs were investigated for the entire family of magma
equations \eqref{eq:magma} in \cite{lowman_dispersive_2013-1}.

\subsection{Outline of this work}
\label{sec:outline-this-work}

The paper is organized as follows.  Periodic traveling wave solutions
to the conduit equation are studied in Sec.~\ref{sec:peri-trav-wave}
both numerically and asymptotically in the weakly nonlinear regime. In
Sec.~\ref{sec:nonl-schr-appr}, we consider weakly nonlinear periodic
wave modulations and include long wave dispersion to derive the NLS
equation. By an appropriate choice of the periodic traveling wave's
wavenumber, both the focusing and defocusing variants of the NLS
equation are possible.  We numerically demonstrate the persistence of
dark and bright modulation solitary wave solutions in the full conduit
equation. In Sec.~\ref{sec:whitham-equations}, we analyze modulated
periodic waves of arbitrary amplitude via the conduit Whitham
modulation equations.  By weakly nonlinear analysis and direct
numerical computation, we determine structural properties of the
Whitham equations including hyperbolicity or ellipticity and genuine
nonlinearity or linear degeneracy. The manuscript is concluded with a
discussion of the implications of this work in Section
\ref{sec:conclusion}.

\section{Periodic traveling wave solutions}
\label{sec:peri-trav-wave}

We seek periodic traveling wave solutions to Eq.~\eqref{eq:conduit} in
the form $A(z,t) = \phi(\theta)$, $\theta=kz-\om t$, $\phi(\theta +
2\pi) = \phi(\theta)$ for $\theta \in \R$. Inserting this ansatz into
Eq.~\eqref{eq:conduit} yields
\begin{equation}\label{eq:TW1}
-\om \phi' + k(\phi^2)' + \om k^2(\phi^2(\phi^{-1}\phi')')' = 0.
\end{equation}
Integrating twice results in
\begin{equation}\label{eq:TW2}
  (\phi')^2 = g(\phi)\equiv -\frac{2}{k^2}\phi - \frac{2}{\om
    k}\phi^2\ln{\phi} +A + B \phi^2, 
\end{equation}
where $A$ and $B$ are real integration constants.

Equation \eqref{eq:TW2} exhibits at most three real roots
\cite{lowman_dispersive_2013-1}. When there are three distinct roots,
a periodic solution oscillates between the largest two. The solution
can be parameterized by three independent variables. Defining the wave
minimum $\phi_0$ according to $\phi_0 = \min_\theta \phi(\theta)$, we
utilize the following physical parametrization
\begin{equation}\label{eq:TW5}
  \begin{array}{rc}
    \text{wavenumber:} & k, \\[3mm]
    \text{wave amplitude:} & \displaystyle a = \max_{\theta \in [0,\pi]}
    \phi(\theta)-\phi_0 , \\ 
    \text{wave mean:} & \displaystyle 
    \phib\equiv\frac{1}{\pi}\int_0^\pi \phi(\theta)\rmd \theta =
    \frac{1}{\pi}\int_{\phi_0}^{\phi_0+a} \frac{\phi\rmd\phi}{\sqrt{g(\phi)}} .
  \end{array}
\end{equation}
The requirement that $\phi$ is $2\pi$-periodic is enforced through
\begin{equation}\label{eq:TW4}
    \pi = \int_0^\pi \rmd\theta =
    \int_{\phi_0}^{\phi_0+a}\frac{\rmd\phi}{\sqrt{g(\phi)}}, 
\end{equation}
where in \eqref{eq:TW5} and \eqref{eq:TW4} we have used the even
symmetry of solutions to eq.~\eqref{eq:TW2}.  Given $(k,a,\phib)$,
the relations \eqref{eq:TW5} and \eqref{eq:TW4} determine the wave
frequency $\om = \om(k,a,\phib)$ and the wave minimum $\phi_0 =
\phi_0(k,a,\phib)$. The extrema requirements $g(\phi_0) =
g(\phi_0+a)=0$ determine $A$ and $B$ from Eq. \eqref{eq:TW2}.

Due to the scaling invariance Eq.~\eqref{eq:scaling1}, the wave mean can
be scaled to unity. This implies that only $\omega(k,a,1)$ and
$\phi_0(k,a,1)$ need be determined.  Then the general cases follow
according to
\begin{align}
  \label{eq:scaling2}
  \om(k,a,\phib) &= \phib^{1/2}\omega\left( \phib^{1/2}k,\phib^{-1}a,1
  \right), \quad
  \phi_0(k,a,\phib) = \phib\phi_0\left(\phib^{1/2}k,\phib^{-1}a,1\right).
\end{align}
We therefore define the unit-mean dispersion and wave solution
according to
\begin{equation}
  \label{eq:3}
  \begin{split}
        \wt(\kt,\at) = \omega(\kt,\at,1), \quad
        \tilde{\phi}(\theta;\kt,\at) &= \phi(\theta; \kt,\at, 1) . 
  \end{split}
\end{equation}
We will use the variables $(\tilde{\phi},\wt,\kt,\at)$ whenever we are
assuming a unit mean solution.

\subsection{Stokes expansion}
\label{sec:stokes-expansion}

\begin{figure}
    \centering
    \includegraphics{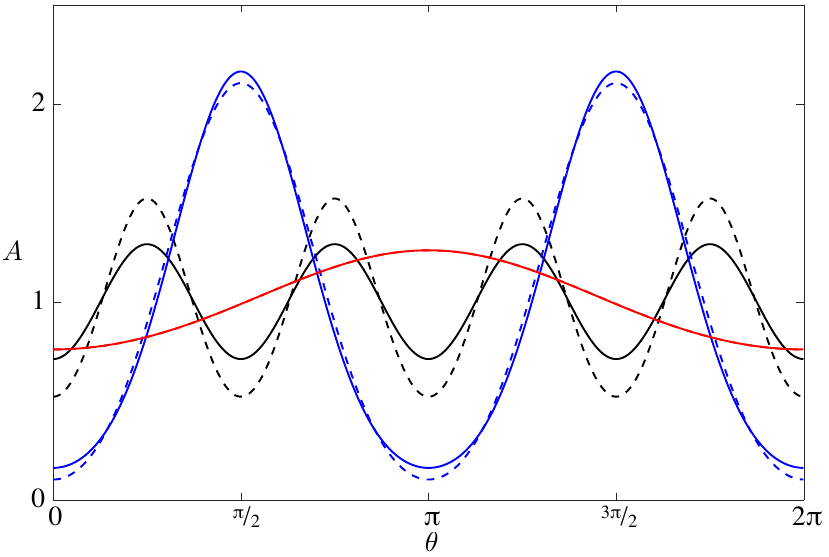}
    \caption{Comparison of the Stokes wave expansion solution (dashed
      lines) to the numerically computed solution (solid lines) for
      three different waves with unit mean. $(k,a)$ is $(1/4,1)$
      (black), $(2,2)$
      (blue), and $(1,0.5)$ (red).}
    \label{fig:stokes_expansion}
\end{figure}

\begin{figure}
  \centering
  \includegraphics{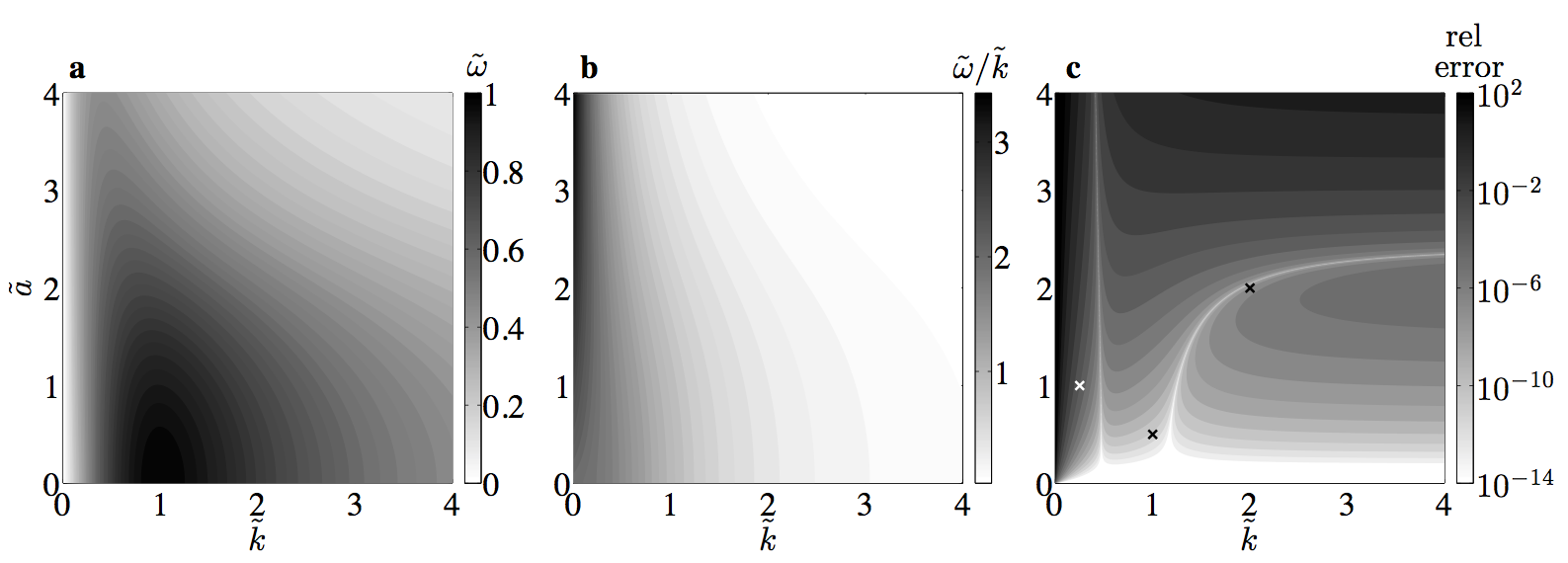}
  \caption{a) Contour plot of numerically computed dispersion
    relation. b) Numerically computed phase velocity. c) Relative
    error between numerically computed dispersion $\wt(\kt,\at)$ and
    approximate dispersion $\wt_0(\kt) +\at^2\wt_2(\kt)$.  Markers
    ($\scriptscriptstyle\pmb{\times}$) correspond to waves plotted in
    Fig.~\ref{fig:stokes_expansion}.}
    \label{fig:dispersion}
\end{figure}

We can obtain approximate periodic traveling wave solutions in the
weakly nonlinear regime via the Stokes wave expansion
\cite{whitham_general_1965}:
\begin{align}
  \tilde{\phi} &= 1+\eps\tilde{\phi}_1 + \eps^2\tilde{\phi}_2 +
  \eps^3\tilde{\phi}_3+\cdots, \\ 
    \wt  &= \wt_0 + \eps^2\wt_2 + \cdots,
\end{align}
where $0 < \eps \ll 1$ is an amplitude scale.  Inserting this ansatz
into Eq. \eqref{eq:TW1}, equating like coefficients in $\eps$, and
enforcing solvability conditions yields the approximate solution
\begin{equation}
  \label{eq:Stokes2}
  \begin{array}{ccccc}
    \tilde{\phi}_1(\theta) = \cos{\theta}, & \wt_0(k) =
    \frac{2\kt}{1+\kt^2}, & 
    \tilde{\phi}_2(\theta) = \frac{1}{6\kt\wt_0}\cos{2\theta},
    &\wt_2(\kt)= \frac{1-8\kt^2}{48\kt(1+\kt^2)},
  \end{array}
\end{equation}
where $\wt_0$ is the unit mean linear dispersion relation \eqref{eq:lineardispersion}.  Setting the
amplitude $\at = 2\eps$, the approximate periodic wave solution is
\begin{align}
  \tilde{\phi}(\theta) &=
  1+\frac{\at}{2}\cos{\theta}+\frac{\at^2(1+\kt^2)}{48
    \kt^2}\cos{2\theta}+ \mathcal{O}(\at^3), \label{eq:SE3a}
  \\
  \wt(\kt,\at) &= \frac{2 \kt}{1+\kt^2} + \at^2\frac{1-8
    \kt^2}{48\kt (1+ \kt^2)} +\Order{\at^3} \label{eq:SE3b} .
\end{align}
In Fig \ref{fig:stokes_expansion}, this solution is compared to
numerically computed periodic waves (numerical methods are described
in Appendix \ref{sec:periodic-solutions}).  The frequency and wave
profile of the Stokes expansion accurately describe some periodic
conduit waves, even for $\mathcal{O}(1)$ amplitudes provided the
wavenumber is appropriately chosen. However, even at moderately small
wavenumbers, the expansion rapidly breaks down. This is quantified in
Fig.~\ref{fig:dispersion}.
Figure \ref{fig:dispersion}(a,b) shows the dispersion and phase
velocity for numerically computed periodic waves, and (c) compares the
full, nonlinear dispersion $\wt(\kt,\at)$ to
$\wt_0(\kt)+\at^2\wt_2(\kt)$. Note the dispersion relation agrees
exceedingly well for $\kt>1$ and $\at\lesssim 1$, but deviates for
larger amplitudes and wavenumbers less than $0.5$.

Of interest is that the approximate solution \eqref{eq:SE3a} can
result in an unphysical, negative conduit cross-sectional area. The
minimum of the approximate solution $\tilde{\phi}(\theta)$ occurs when
$\theta = \pi$. Equating the minimum to zero, we find that physical,
positive values for approximate $\tilde{\phi}$ are restricted to $\at
< a_0$, where $a_0 > 4(3 - \sqrt{6}) \approx 2.20$, which is well
beyond our assumption of small amplitude $0 < \at \ll 1$.

\section{Weakly nonlinear, dispersive modulations}
\label{sec:nonl-schr-appr}

\begin{figure}
  \centering
  \includegraphics{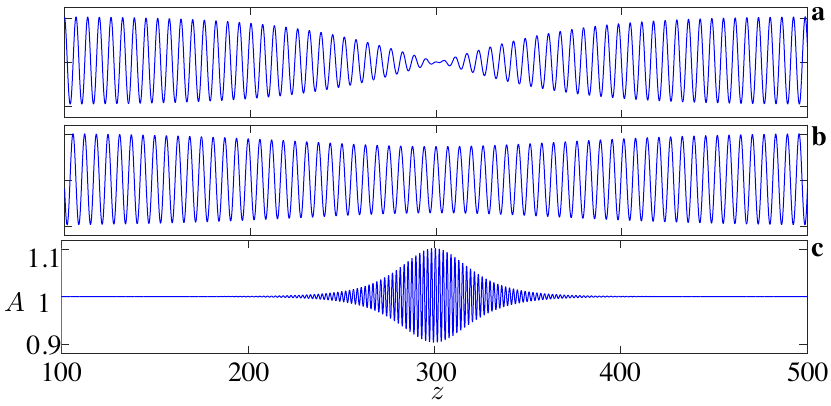}
  \caption{Examples of (a) black, (b) gray, and (c) bright envelope
    solitons from the NLS approximation. These approximate solutions
    were found to persist for the conduit equation over long time ($t
    = \mathcal{O}(1/\eps^2)$), provided the modulated wave was of
    amplitude $\at = 2\eps\lessapprox 0.4$.  The pictured dark (bright)
    solitons are on a background wave with $\eps=0.1$ and $\kt=1$
    ($\kt=3$). The gray soliton exhibits the phase jump $\alpha =
    \pi/4$.}
  \label{fig:envelope_solis}
\end{figure}

The aim of this section is to describe wave modulation in the weakly
nonlinear regime. The approximate NLS equation is found using multiple
scales in Appendix \ref{sec:nonl-schr-deriv} by seeking a solution in
the form 
\begin{equation}
  \label{eq:NLSfinal}
  \begin{split}
    A(z,t) &= 1+\eps\left[\sqrt{n}B e^{i\theta}+c.c.\right]+
    \eps^2\left[\frac{nB^2}{3\kt\wt_0}e^{2i\theta}+c.c.+ M
    \right]+ \mathcal{O}(\eps^3),\\ 
    M &= \frac{(3\kt-1)(1+\kt^2)}{\kt^2(\kt^2+3)}n\abs{B}^2, \quad
    n(\kt) = \frac{3+5\kt^2+8\kt^4}{3\kt(\kt^2+1)(\kt^2+3)} 
  \end{split}
\end{equation}
where $c.c.$ denotes complex conjugate and $\eps$ is an amplitude
scale. By introducing the standard, scaled coordinate system
$$ \tau = \eps^2t, \quad \zeta =
\frac{\eps}{\sqrt{\abs{\wt_0''}}}(z-\wt_0't) ,$$ 
we obtain the NLS equation for the complex envelope $B(\zeta,\tau)$
\begin{equation}
  \label{eq:NLS}
  iB_\tau + \frac{\sigma}{2}B_{\zeta\zeta} + \abs{B}^2B = 0,
\end{equation}
where $\sigma = \sgn \wt_0''(\kt)$ denotes the dispersion curvature.  Since
$$ \wt_0''(\kt)=\frac{4\kt(\kt^2-3)}{(1+\kt^2)^3},$$
the NLS equation \eqref{eq:NLS} is defocusing when $0<\kt<\sqrt{3}$,
and focusing for $\kt>\sqrt{3}$.  This result effectively splits
periodic wave solutions of the conduit equation into two regimes.  For
the defocusing case, weakly nonlinear periodic waves are
modulationally stable, and dark envelope solitons are predicted,
which, when combined with the ansatz \eqref{eq:NLSfinal}, take the
form (see, e.g., \cite{ablowitz_nonlinear_2011})
\begin{align}
    B(\zeta,\tau) &= \tanh(\zeta) e^{i\tau}, \tag{black}
  \nonumber \\
  B(\zeta,\tau) &= e^{2i\tau+i\psi_0}\left[\cos\alpha+i\sin\alpha
    \tanh[\sin\alpha\, (\zeta-2\cos\alpha\,
    \tau-\zeta_0)]\right], \tag{gray}
\end{align}
with arbitrary, real constants $\zeta_0$, $\psi_0$, and
$\alpha$, where $\alpha$ is the phase jump across the gray soliton.
Examples of modulated periodic waves in this regime are
shown in Figs. \ref{fig:envelope_solis}(a,b).
For the focusing case, periodic waves are modulationally unstable
\cite{zakharov_modulation_2009}. Bright envelope solitons are
predicted, which have the form
\begin{align}
  B(\zeta,\tau) &=\sech(\zeta-\zeta_0)
  e^{-i\Theta},\quad \Theta =-2\tau+\Theta_0,\tag{bright}
\end{align}
where $\zeta_0$ and $\Theta_0$ are arbitrary, real constants.
An example of a bright envelope soliton is shown in
Fig. \ref{fig:envelope_solis}(a).

\begin{figure}
  \includegraphics{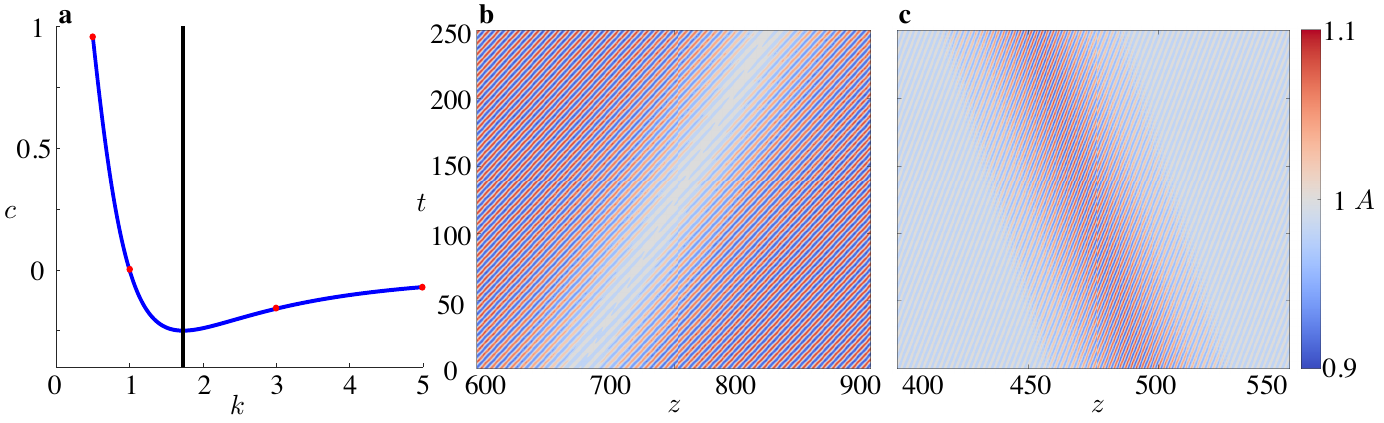}
  \caption{(a) Comparison of envelope soliton velocities extracted from
    numerical simulation (dots) to the linear group velocity (solid
    line). The vertical line denotes the crossover from defocusing to
    focusing NLS. (b,c) Contour plots of dark (b) and bright (c) envelope
    solitons. The dark (bright) soliton modulates a periodic wave with
    $\eps=0.05$ and $\kt=0.5$ ($\kt=3$).}
  \label{fig:envelope_soli_contours}
\end{figure}

We find that approximate bright and dark envelope solitons persist to
times $t=\Order{1/\eps^2}$ by direct numerical simulation of the
conduit equation \eqref{eq:conduit} with envelope soliton initial
data.  Appendix \ref{sec:time-stepping} describes our numerical
time-stepping method.  The evolution of a black and a bright envelope
soliton is depicted in the contour plots of
Fig.~\ref{fig:envelope_soli_contours}.  While the black soliton
Fig~\ref{fig:envelope_soli_contours}(a) propagates forward in time,
the bright soliton \ref{fig:envelope_soli_contours}(b) has a negative
velocity.

The velocity of both the dark and bright approximate envelope solitons
are, to leading order in $\eps$, the group velocity
Eq.~\eqref{eq:lineardispersion}.  We find excellent agreement between
the predicted group velocity and the velocity extracted from numerical
simulation as shown in Fig.~\ref{fig:envelope_soli_contours}(a).  Envelope
solitons, dark or bright, for $\kt > 1$ exhibit negative velocity.

For simulation times sufficiently large, $t \gg 1/\eps^2$, envelope
solitons emit radiation. The dark soliton decays and the bright
soliton decomposes into smaller wave packets.  We also find that
numerical simulation of envelope solitons modulating periodic waves of
sufficiently large amplitude ($\eps \gtrapprox 0.1$) breakdown almost
immediately, although some dark envelope solitons appear to asymptote
to a stable but different envelope form.  Due to the breakdown of the
NLS approximation at finite amplitudes, we require a more accurate
description of nonlinear wave modulations in the moderate to large
amplitude regime.

\section{Whitham equations}
\label{sec:whitham-equations}

In order to describe modulated, large amplitude periodic waves, we
appeal to the Whitham modulation equations.  Whitham's original
formulation invoked averaged conservation laws
\cite{whitham_non-linear_1965}, later shown to be equivalent to a
perturbative, multiple-scales reduction \cite{luke_perturbation_1966}.
For completeness, we have implemented the multiple-scales reduction in
Appendix \ref{sec:whith-equat-deriv}.  For this, we seek modulations
of an arbitrary amplitude, $2\pi$-periodic, traveling wave solution
$\phi$ to eq.~\eqref{eq:conduit} (see Sec.~\ref{sec:peri-trav-wave}).
Since we
will incorporate only the leading order Whitham equations, the large
parameter $1/\eps$ here corresponds to the time scale of their
validity.  Note that this is shorter than the $\mathcal{O}(1/\eps^2)$
time scale of the NLS equation \eqref{eq:NLS}.  Then we arrive at the
Whitham equations (Appendix \ref{sec:whith-equat-deriv})
\begin{equation}
  \label{eq:whitham1}
  \begin{split}
    \left(\overline{\phi}\right)_t +  \left(\overline{\phi^2}
      -2k\om\overline{g(\phi)}\right)_z &=0, \\ 
    \left(\overline{\frac{1}{\phi}} +
      k^2\overline{\frac{g(\phi)}{\phi^2}} \right)_t
    -2\left(\overline{\ln \phi}\right)_z &=0, \\ 
    k_t + \om_z &= 0 ,
  \end{split}
\end{equation}
where we recall the defining ordinary differential equation (ODE) $\phi'^2 =
g(\phi)$ \eqref{eq:TW2} and utilize the notation
$$
\overline{f} = \frac{1}{2\pi} \int_0^{2\pi}
f(\theta)\operatorname{d}\theta .
$$

It is convenient to express the Whitham equations in the form
\begin{equation}\label{eq:whitham2}
    \begin{array}{ccc}
        \Pcal_t+\Qcal_z = 0,  &  
        \Pcal =  \begin{bmatrix}
                    \phib \\ I_1 \\ k
                 \end{bmatrix},  &
        \Qcal =  \begin{bmatrix}
                    I_2 \\ I_3 \\ \om 
                 \end{bmatrix},
    \end{array}
\end{equation}
where we have introduced the averaging integrals
\begin{equation}
  \label{eq:whithamIntegrals}
  \begin{array}{ccc}
    I_1 = \overline{\phi^{-1}} + k^2\overline{g(\phi)/\phi^2}, &
    I_2 = \overline{\phi^2} - 2k\om\overline{g(\phi)}, &
    I_3 = -2\overline{\ln{\phi}}.
  \end{array}
\end{equation}
Furthermore, we can expand the density $\Pcal$ and flux $\Qcal$ in
terms of the modulation variables $\mathbf{q}=(k,a,\phib)^T$ to obtain
the quasi-linear form of the Whitham equations
\begin{equation}
  \label{eq:whitham3}
  \mathbf{q}_t + \Acal \mathbf{q}_z = 0,
\end{equation}
where
\begin{equation}
  \label{eq:7}
  \begin{split}
    \Acal &=\left(\dfrac{\prt\Pcal}{\prt\mathbf{q}}\right)^{-1}
    \dfrac{\prt\Qcal}{\prt\mathbf{q}} \\
    &=  
    \begin{bmatrix}
      \om_k & \om_a & \om_{\phib} \\
      \frac{I_{3,k}  -I_{1,k}\om_k -I_{2,k}I_{1,\phib}}{I_{1,a}} &
      \frac{I_{3,a}  -I_{1,k}\om_a -I_{2,a}I_{1,\phib}}{I_{1,a}} &
      \frac{I_{3,\phib}-I_{1,k}\om_{\phib}-I_{2,\phib}I_{1,\phib}}{I_{1,a}}
      \\ 
      I_{2,k} & I_{2,a} & I_{2,\phib}
    \end{bmatrix} .
  \end{split}
\end{equation}
This non-conservative form of the Whitham equations is only valid
where the matrix $\partial \mathcal{P}/\partial \mathbf{q}$ is
invertible.

The scaling invariance \eqref{eq:scaling1} can be used so that the
dependence on $\phib$ in the Whitham equations is explicit and the
averaging integrals need be computed only over the scaled variables
$\kt$ and $\at$.  Then the integrals \eqref{eq:whithamIntegrals} can
be written
\begin{equation}
  \label{eq:8}
    \begin{array}{ccc}
        I_1 = \frac{1}{\phib}\tilde{I_1}, & 
        I_2 = \phib^2\tilde{I_2}, &
        I_3 = \tilde{I_3}-2\ln{\phib},
    \end{array}
\end{equation}
where $\tilde{I}_i = \tilde{I}_i(\kt,\at)$, $i=1,2,3$. Therefore,
computation of the averaging integrals is only required for
$(\kt,\at)$. The Whitham equations in the scaled variables $\qt =
(\kt,\at,\phib)$ are
\begin{equation}\label{eq:whitham4}
    \begin{array}{ccc}
        \qt_t + \tilde{\Acal}\qt_z = 0, &  
        \tilde{\Acal} =
        \left(\dfrac{\prt\mathbf{q}}{\prt\qt}\right)^{-1}
        \Acal\dfrac{\prt\mathbf{q}}{\prt\qt},  
        & 
        \dfrac{\prt\mathbf{q}}{\prt\qt} = 
        \begin{bmatrix}
            \phib^{-1/2} &   0   & -\frac{1}{2}\phib^{-3/2}\kt \\
                 0       & \phib & \at                         \\
                 0       &   0   & 1
        \end{bmatrix}.
    \end{array}
\end{equation}

We will be interested in structural properties of the Whitham
equations such as hyperbolicity (strict or non-strict),
ellipticity, and genuine nonlinearity.  All of these criteria rely on
the eigenvalues and eigenvectors of the Whitham equations that satisfy
\begin{equation}
  \label{eq:4}
  ( \Acal - c I) \mathbf{r} = 0 .
\end{equation}
In general, we expect three eigenpairs
$\{(c_j,\mathbf{r}_j)\}_{j=1}^3$ with either all real eigenvalues $c_1
\le c_2 \le c_3$, where the Whitham equations are hyperbolic.  If the
inequalities are strict, then the Whitham equations are strictly
hyperbolic.  Or, in the case of one real and two complex conjugate
eigenvalues, the Whitham equations are elliptic.

The coefficient matrix $\Acalt$ is a similarity transformation of
$\Acal$ so its eigenvalues are the same.  $\Acalt$ exhibits the
following property
\begin{equation}
  \label{eq:37}
  \Acalt(\kt,\at,\phib) =
  \begin{bmatrix}
    1 & 0 & 0 \\
    0 & 1 & 0 \\
    0 & 0 & \phib
  \end{bmatrix} \Acalt(\kt,\at,1)
  \begin{bmatrix}
    \phib & 0 & 0 \\
    0 & \phib & 0 \\
    0 & 0 & 1
  \end{bmatrix} ,
\end{equation}
which can be used to show
\begin{equation}
  \label{eq:49}
  c(\kt,\at,\phib) = \phib c(\kt,\at,1), \quad
  \mathbf{r}(\kt,\at,\phib) = \begin{bmatrix}
    \phib^{-1} & 0 & 0 \\
    0 & \phib^{-1} & 0 \\
    0 & 0 & 1
  \end{bmatrix} \mathbf{r}(\kt,\at,1) .
\end{equation}
Therefore, the hyperbolicity/ellipticity of the Whitham equations is
independent of the mean $\phib$.  We define the unit mean eigenvalues
$\tilde{c}$ and eigenvectors $\rt$ according to
\begin{equation}
  \label{eq:9}
  \ct(\kt,\at) = c(\kt,\at,1), \quad \rt(\kt,\at) =
  \mathbf{r}(\kt,\at,1) .
\end{equation}

Utilizing the identities in \eqref{eq:49}, we find that the quantity
\begin{equation}
  \label{eq:55}
  \mu \equiv \nabla_{\qt} c(\kt,\at,\phib) \cdot \mathbf{r}(\kt,\at,\phib) =
  \begin{bmatrix}
    \ct_{\kt} \\ \ct_{\at} \\ \ct
  \end{bmatrix} \cdot
  \rt 
\end{equation}
is independent of $\phib$, $\mu = \mu(\kt,\at)$.  If $\mu \ne 0$, then
the Whitham equations are genuinely nonlinear
\cite{whitham_linear_1974}.  For those values of $\kt$ and $\at$ where
$\mu = 0$, the Whitham equations are linearly degenerate.  The sign
definiteness of $\mu$ corresponds to a monotonicity condition that is
required for the existence of simple wave solutions to the Whitham
equations, of particular importance for the study of DSWs
\cite{el_dispersive_2016}.

\subsection{Weakly nonlinear regime}
\label{sec:weakly-nonl-regime}

We now consider Eq.~\eqref{eq:7} in the small $a$ regime by inserting
the Stokes expansion \eqref{eq:SE3a}, \eqref{eq:SE3b}, yielding
\begin{equation}\label{eq:smallAmp2}
  \begin{array}{cc}
    \Acal = 
    \begin{bmatrix}
      \om_{0,k}             & 2a\om_2   & \om_{0,\phib} \\
      \frac{a}{2}\om_{0,kk} & \om_{0,k} & \frac{a}{2}\frac{4(1+\phib
        k^2+3\phib^2k^4 + \phib^3k^6)}{(1+\phib k^2)^3} \\ 
      0                     & 2a\frac{1-3\phib k^2}{8(1+\phib k^2)}
      & 2\phib 
    \end{bmatrix} + \mathcal{O}(a^2) .
  \end{array}
\end{equation}
We directly compute the characteristic velocities and, via
\eqref{eq:49}, evaluate at unit mean $\phib = 1$
\begin{align}
  \label{eq:smallAmpSpeeds}
  \tilde{c}_1 &= \wt_{0,\kt} - \frac{\at}{4}\sqrt{-n\wt_{0,\kt\kt}} +
  \Order{\at^2}, \nonumber\\
  \tilde{c}_2 &= \wt_{0,\kt} + \frac{\at}{4}\sqrt{-n\wt_{0,\kt\kt}} +
  \Order{\at^2}, \\
  \tilde{c}_3 &= 2 + \Order{\at^2}, \nonumber
\end{align}
where $n=n(\kt)$ is from Eq.~\eqref{eq:NLSfinal}.  We observe that
complex characteristic velocities occur precisely when the NLS
equation \eqref{eq:NLS} is in the focusing regime, i.e., when $\kt >
\sqrt{3}$.  This is to be expected \cite{whitham_linear_1974}.  The
requirement $-n\wt_{0,\kt\kt} > 0$ for modulational stability is
sometimes referred to as the Benjamin-Feir-Lighthill criterion
\cite{zakharov_modulation_2009}.  Note that we must use $-n(\kt)$, as
opposed to $\wt_2(\kt)$ from the Stokes expansion \eqref{eq:Stokes2},
in the criterion because of the generation of a mean term
(cf.~\cite{whitham_linear_1974}).

Because the linear group velocity \eqref{eq:linearvelocities} is
always less than the linear, long wave velocity $2$, the weakly
nonlinear approximation of the Whitham equations imply strict
hyperbolicity $\tilde{c}_1 < \tilde{c}_2 < \tilde{c}_3$ for $\at > 0$,
$\kt^2<3$.

Next, we determine the approximate eigenvectors $\rt_i$ associated
with the approximate eigenvalues \eqref{eq:smallAmpSpeeds} using
standard asymptotics of eigenvalues and eigenvalues (see, e.g.,
\cite{hinch_perturbation_1991}).  These approximate results are used to compute
$\mu_i$, $i = 1,2,3$ \eqref{eq:55}.  The expressions are cumbersome
but there are two noteworthy findings.  We find that $\mu_1 = 0$ when
\begin{equation}
  \label{eq:10}
  \at = 6 \sqrt{\frac{2}{5}} 3^{3/4}\left ( \sqrt{3} - \kt \right )^{3/2} +
  \mathcal{O} \left (\left (\sqrt{3}-\kt \right)^{5/2} \right ), \quad
  0 < \sqrt{3} - \kt \ll 1 .
\end{equation}
Along the curve \eqref{eq:10}, the weakly nonlinear Whitham equations
are linearly degenerate in the first characteristic field.

We also find linear degeneracy in the second characteristic field:
$\mu_2 = 0$ when $\at = 18 \kt^2 + \Order{\kt^4}$, $0 < \kt \ll 1$.
But the weakly nonlinear Whitham equations are nonstrictly hyperbolic,
$\ct_2 = \ct_3$, when $\at = 12\kt^2 + \Order{\kt^4}$, $0 < \kt \ll
1$.  Because nonstrict hyperbolicity implies linear degeneracy
\cite{dafermos_hyperbolic_2009}, we have apparently obtained a
contradiction.  We argue that this is due to the asymptotic
approximations made and the poor accuracy of the approximate periodic
wave solution afforded by the Stokes expansion \eqref{eq:SE3a},
\eqref{eq:SE3b} for small $\kt$ (cf.~Figs.~\ref{fig:stokes_expansion},
\ref{fig:dispersion}).


\subsection{Large amplitude regime}
\label{sec:large-ampl-regime}

We now investigate modulations of large amplitude, periodic waves by
direct computation of the Whitham equations.  For this, we examine the
Whitham equations in the form \eqref{eq:whitham4}, so that the
dependence on $\phib$ is explicit.  We numerically compute periodic
solutions $\tilde{\phi}$ and the corresponding dispersion
$\wt(\kt,\at)$ and unit-mean averaging integrals
$\{\tilde{I}_i(\kt,\at)\}_{i=1}^3$ for the equispaced, discrete values
$(\kt_j,\at_l)$, $\kt_j = j\Delta$, $\at_l = l\Delta$, $j,l = 1, 2,
\ldots, N$.  We chose $N = 4000$, $\Delta = 0.001$ so that $\kt_N =
\at_N = 4$.  Derivatives of $\wt$ and $\tilde{I}_j$ with respect to
$\kt$ and $\at$, required in \eqref{eq:whitham3}, are estimated with
sixth order finite differences, yielding a numerical approximation of
the coefficient matrix $\Acalt$ on the discrete grid.

\begin{figure}
  \centering
  \includegraphics{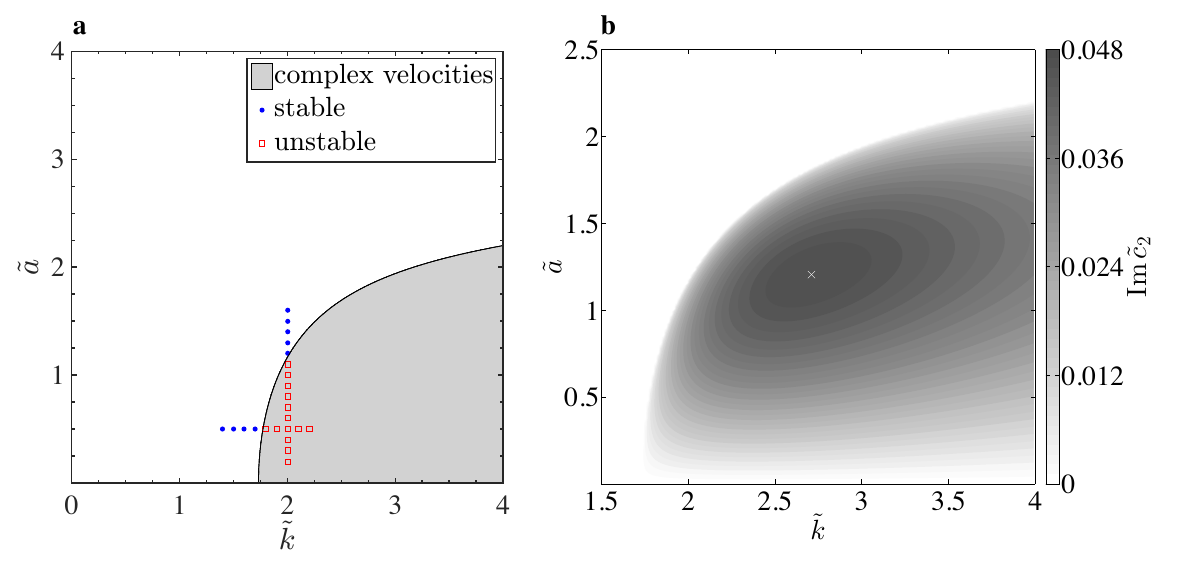}
  \caption{a) Elliptic (gray) and hyperbolic (white) parameter regimes
    for the Whitham equations corresponding to complex or real
    characteristic velocities, respectively.  Stable (dots) and
    unstable (squares) periodic waves according to direct numerical
    simulation of the conduit equation. b) Contour plot of the
    imaginary part of the characteristic velocity $\tilde{c}_2$, the
    MI growth rate.  The maximum, $0.04795$, occurs for
    $(\tilde{k},\tilde{a}) = (2.711,1.204)$.}
  \label{fig:mi_regime}
\end{figure}

Using our direct computation of the coefficient matrix $\Acalt$, we
determine its eigenvalues $\{\tilde{c}_i(\kt,\at)\}_{i=1}^3$ and plot
in Fig.~\ref{fig:mi_regime}a the region in the $\kt$-$\at$ plane where
the Whitham equations are hyperbolic or elliptic.  As shown by our
weakly nonlinear analysis \eqref{eq:smallAmpSpeeds}, the elliptic
region appears for $\kt > \sqrt{3}$, independent of $\at$ for small
$\at$.  But our computations show that the region depends strongly on
the wave amplitude.

\begin{figure}
  \centering
  \includegraphics{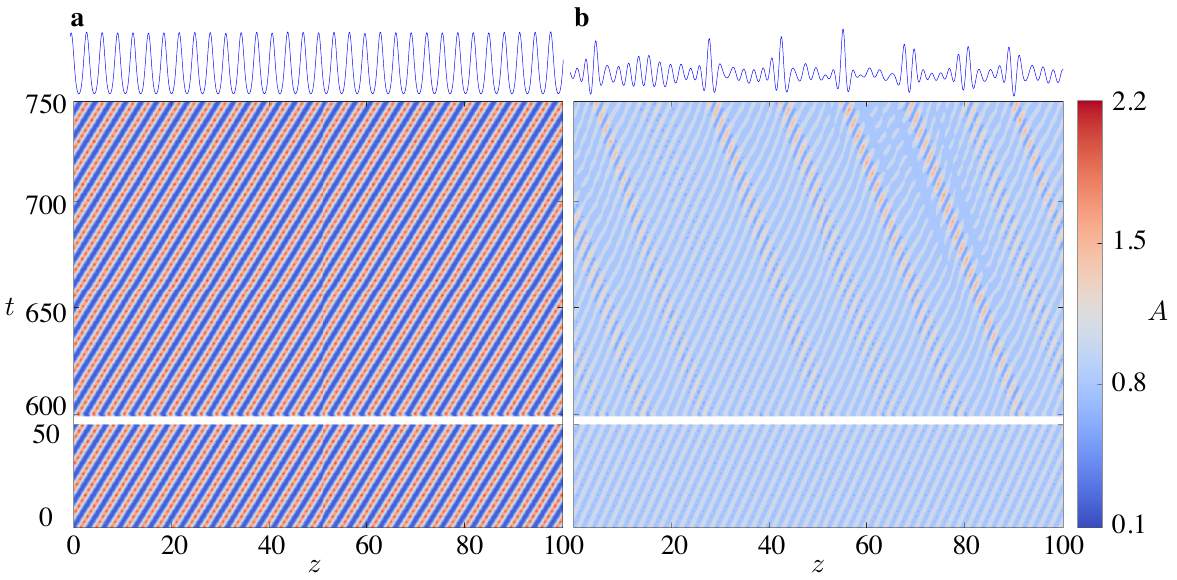}
  \caption{Numerical evolution of perturbed periodic wave solutions in
    the conduit equation. a) Modulationally stable case:
    $(\kt,\at)=(2,2)$. b) Modulationally unstable case:
    $(\kt,\at)=(3,0.5)$. Top: the respective cases at
    the final time $t=750$.}
  \label{fig:exm_periodic_waves}
\end{figure}

As noted earlier, ellipticity of the Whitham equations implies
modulational instability of the periodic traveling wave
\cite{zakharov_modulation_2009}.  In agreement with our weakly
nonlinear analysis \eqref{eq:smallAmpSpeeds}, we find that in the
elliptic region, $\ct_1 = \ct_2^*$ ($^*$ denotes complex conjugation)
and $\ct_3 \in \mathbb{R}$.  We confirm the hyperbolic/elliptic
boundary by direct numerical simulation of the conduit equation
\eqref{eq:conduit} with slightly perturbed, periodic initial data.
Random, smooth noise (band-limited to wavenumber 512) of magnitude
$\Order{10^{-3}}$ was added to a periodic traveling wave initial
condition on a domain of over 100 spatial periods.  This initial data
was evolved either 100 temporal periods or to $t = 500$, whichever was
longer. Some waves, especially those in the small-amplitude regime,
were evolved for even longer time periods. The modulational
(in)stability of several of these runs are shown in
Fig.~\ref{fig:mi_regime}(a). We find excellent agreement with the
MI predictions from Whitham theory. The long-time evolution of two
particular waves are shown in Fig.~\ref{fig:exm_periodic_waves},
showing both a stable and an unstable case.

A periodic traveling wave solution of the conduit equation
\eqref{eq:conduit} corresponds to a constant solution $\qt(z,t) =
\qt_0$ of the Whitham equations \eqref{eq:whitham4}.  If we consider
the stability of this solution by linearizing the Whitham equations
according to $\qt(z,t) = \qt_0 + \mathbf{b} e^{i \kappa z + \sigma t}$,
$|\mathbf{b}| \ll 1$, we obtain the growth rates
\begin{equation}
  \label{eq:11}
  \sigma_i = \kappa \,\mathrm{Im} (c_i),
\end{equation}
for each component of the perturbation in the eigenvector basis of
$\Acalt$.  The physical growth rate requires knowledge of the
wavenumber $\kappa$.  Because the Whitham equations are quasi-linear,
first order equations, any wavenumber is permissible (determined by
the initial data), suggesting that the physical growth rate
\eqref{eq:11} is unbounded.  In practice, there is a dominant
wavenumber associated with the instability that is determined by
higher order effects, which in this case is dispersion.  The NLS
equation \eqref{eq:NLS} resolves this feature in the weakly nonlinear
regime but we are interested in large amplitude modulations.  We
therefore identify the imaginary part of the characteristic velocity
$\tilde{c}_2$ as a proxy for the growth rate of the instability and
observe in Fig.~\ref{fig:mi_regime}b that there is a maximum of
$\mathrm{Im}(\tilde{c}_2)$ for unit-mean periodic waves that occurs
for the wave parameters $(\tilde{k},\tilde{a}) = (2.711,1.204)$.  We
confirm that these parameters do indeed approximately correspond to a
maximally unstable periodic wave by performing numerical simulations
of the conduit equation \eqref{eq:conduit} with initially perturbed
periodic traveling waves, using the same process used when determining modulational (in)stability. 
The envelopes of these waves were extracted for each time step and then compared to the envelope of the initial condition, giving a deviation from the expected periodic wave evolution. 
The growth rate was calculated by fitting an exponential to this deviation.
From these numerics, the maximally unstable parameters are closer to $(\kt,\at) = (2.7,1.35)$ than the expected $(\kt,\at)=(2.7,1.2)$. 
The maximal growth rate for these parameters was found to be $0.0457$, which is within $5\%$ of the maximal growth rate found via the Whitham equations.
 Thus the $\kappa$ does not drastically affect the maximal growth rates.

\begin{figure}
  \centering
  \includegraphics{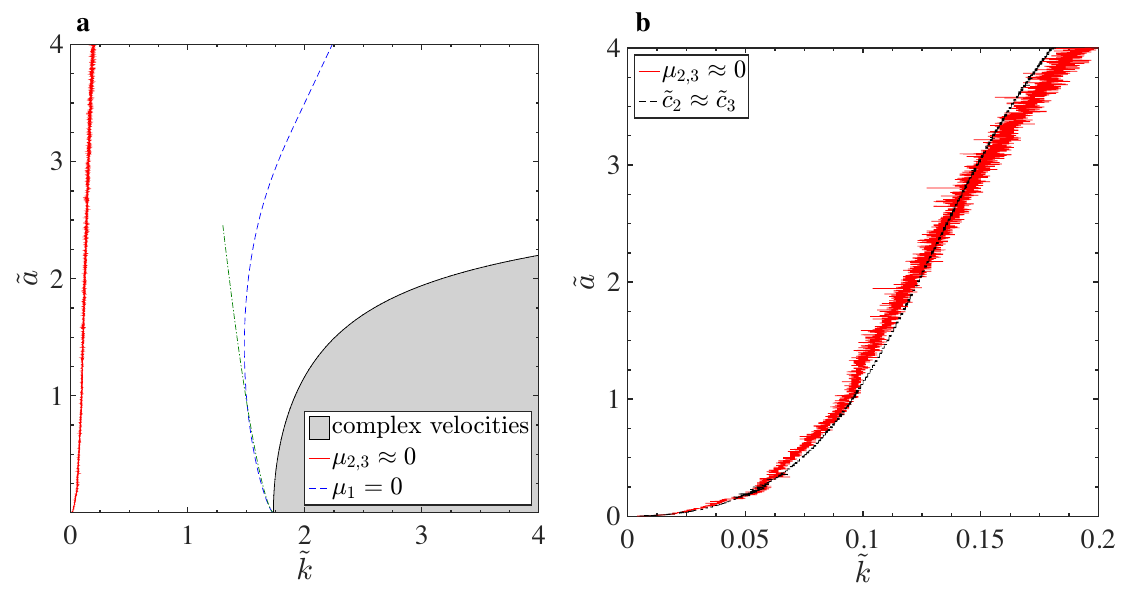}
  \caption{a) Loss of genuine nonlinearity in the Whitham equations.
    The curves correspond to regions in the $\kt$-$\at$ plane where
    the computed quantities $\mu_{2,3}$ (solid) or $\mu_1$ (dashed)
    change sign.  To the right of the solid (dashed) curve, $\mu_{3} >
    0$, $\mu_2 < 0$ ($\mu_{1} > 0$).  The dash-dotted curve
    corresponds to the prediction $\mu_1 = 0$ from a weakly nonlinear
    analysis. The elliptic region from Fig.~\ref{fig:mi_regime} is
    also depicted (gray).  b) Zoom-in of the small $\kt$ region of a)
    where $\mu_{2,3} \approx 0$ (red) approximately corresponds to
    the largest $\kt$, to the left of which $|\ct_3-\ct_2|<10^{-5}$
    (black), i.e., approaching non-strict hyperbolicity.}
  \label{fig:monotonicity}
\end{figure}

Next, we compute the quantities $\{\mu_i(\kt,\at)\}_{i=1}^3$ from
eq.~\eqref{eq:55} on the discrete grid $\{(\kt_j,\at_l)\}_{j,l=1}^N$
using sixth order finite differencing.  The results are depicted in
Fig.~\ref{fig:monotonicity}a where the curves correspond to the
largest value of $\kt$, given $\at$, such that $\mu$ changes sign.
The curve where $\mu_1$ changes sign bifurcates from the edge of the
elliptic region at the point $(\kt,\at) = (\sqrt{3},0)$, agreeing with
the weakly nonlinear result \eqref{eq:10} for sufficiently small
$\at$.  The curve where both $\mu_{2,3}$ change sign apparently
bifurcates from $(0,0)$ and occurs for small $\kt$.  These results
demonstrate that the Whitham equations lack genuine nonlinearity when
considered in the whole of the hyperbolic region.

In order to understand the small $\kt$ results better, we show in
Fig.~\ref{fig:monotonicity}b a zoom-in of this region.  The accurate
determination of the loss of genuine nonlinearity in this region is
numerically challenging because the characteristic velocities
$\ct_{2,3}$ get very close to one another.  A more numerically stable
calculation is shown by the black dashed curve in
Fig.~\ref{fig:monotonicity}b where, for each $\at$, it corresponds to
the largest $\kt$ at which $|\ct_3 - \ct_2| < 10^{-5}$.  For
parameters to the left of this curve, the characteristic velocities
remain very close to one another. It is well-known that, for example,
in the KdV Whitham equations, the characteristic velocities get
exponentially close to one another yet remain distinct in the small
wavenumber regime
\cite{levermore_hyperbolic_1988,gurevich_nonlinear_1990}.  Because
non-strict hyperbolicity implies loss of genuine nonlinearity
\cite{dafermos_hyperbolic_2009}, the proximity of $\ct_2$ and $\ct_3$
may be affecting the numerical results.  It remains to definitively
determine if the Whitham equations lose strict hyperbolicity and/or
genuine nonlinearity in the small $\kt$ regime. Note that the curve
for which $\mu_1 = 0$ in Fig.~\ref{fig:monotonicity}a occurs in a
strictly hyperbolic region.

\section{Discussion/Conclusion}
\label{sec:conclusion}

Our study of the structural properties of the conduit Whitham
equations sheds some light on recent theoretical and experimental
studies of dispersive shock waves.  The DSW fitting method allows one
to determine a dispersive shock's harmonic and soliton edge speeds,
even for non-integrable systems \cite{el_resolution_2005}.  However,
the method is known to break down when the Whitham equations lose
genuine nonlinearity in the second characteristic field
\cite{lowman_dispersive_2013-1,el_dispersive_2016}.  It was observed
in \cite{lowman_dispersive_2013-1} that the fitting method failed to
accurately predict conduit DSW soliton edge speeds for sufficiently
large jump heights.  Our results here suggest that this could be due
to the loss of strict hyperbolicity and/or genuine nonlinearity in the
small wavenumber (soliton train) regime.
    
In addition to the hyperbolic modulation regime where DSWs, and dark
envelope solitons are the prominent coherent structures, we have found
an elliptic regime where the periodic wave breaks up into coherent
wavepackets or bright envelope solitons.  The accessibility of both
hyperbolic and elliptic modulation regimes in one system motivates
further study of each and the transition between the two.  One
potential future, novel direction is to explore the possibility of
creating a soliton gas \cite{el_kinetic_2005}.

It remains to generate a periodic wave from an initially constant
conduit and explore its properties experimentally. 
Accurate control of wavebreaking has been achieved by
slow modulation of the conduit area from a boundary
\cite{maiden_observation_2016}.  One possibility is to utilize simple
wave solutions of the Whitham equations to efficiently and smoothly
transition between a constant conduit $\at = 0$ to a periodic conduit
$\at > 0$.  This also suggests the theoretical and experimental
exploration of Riemann problems, step initial data, for the Whitham
equations themselves. Our determination of linearly degenerate curves
will inform the ability to construct simple waves connecting two
generic wave states.
    
Although our direct numerical simulations suggest that envelope
solitons are long-lived, their physical realization may be challenging.
Existing laboratory studies of viscous fluid conduits implement
control of the conduit interface by varying the injected flow rate
through a nozzle at the bottom of a fluid column.  This allows for the
creation of waves with positive (upward) propagation velocities.
While some dark solitons have a positive velocity, all bright envelope
solitons propagate with negative velocity.  Additionally, it has been
found to be difficult to image small amplitude excitations so that the
observation of weakly nonlinear dynamics may be challenging.
Nevertheless, the experimental creation of periodic conduit waves and
envelope solitons is possible. 
  
We have shown that the non-convexity of the conduit linear dispersion
relation leads to the existence of elliptic Whitham equations and
modulational instability.  This is just one possible implication of
non-convex dispersion in dispersive hydrodynamics.  We note that
non-convex dispersion in other, higher order equations, has also been
found to give rise to a resonance between the DSW soliton edge and
linear waves, leading to the generation of radiating DSWs
\cite{conforti_resonant_2014,el_radiating_2016,sprenger_shock_2016}.

This study has identified and categorized modulations of periodic
traveling waves for the conduit equation \eqref{eq:conduit}. These
findings, along with previous theoretical and experimental studies of
solitons and DSWs imply that the viscous fluid conduit system is an
accessible environment in which to investigate rich and diverse
nonlinear wave phenomena.
%
\appendix
\section{Numerical Methods}

\subsection{Periodic solutions}
\label{sec:periodic-solutions}

We compute unit-mean conduit periodic traveling wave solutions
$\pt(\theta)$ for specified $(\kt,\at)$ with a Newton-GMRES
iterative method \cite{kelley_iterative_1995} on the first integral of
equation \eqref{eq:TW1}
\begin{equation}
  \label{eq:5}
  A + \wt \pt - \kt \pt^2 - \wt \kt^2 \pt \pt'' + \wt \kt^2
  (\pt')^2 = 0 ,
\end{equation}
where $A \in \mathbb{R}$ is an integration constant.  We use a
spectral method to compute the unit-mean cosine series representation
$\pt(\theta) = 1 + \sum_{n=1}^N 2 a_n \cos n\theta$.  Equation
\eqref{eq:5} is discretized in spectral space $\{a_n\}_{n=1}^N$ with
the fast and accurate computation of derivatives achieved via fast
cosine transforms (DCT II in \cite{wang_discretew_1985}).  The
projection of \eqref{eq:5} onto constants determines $A$, which we do
not require because of our imposition of unit mean.  Projection of
Eq.~\eqref{eq:5} onto $\cos(n\theta)$ for $n=1,\ldots, N$ yields $N$
equations for the $N+1$ unknowns $((a_n)_{n=1}^N,\wt)$.  The amplitude
constraint $\pt(\pi)-\pt(0) = -4\sum_{n\,\mathrm{odd}} a_n = \at$
closes the system of equations.  We precondition the spectral
equations by dividing each by the sum of linear coefficients, shifted
by $2k+1$, i.e., by $\wt + n^2 \wt \kt^2 + 1$.  The accurate
resolution of each solution is maintained by achieving an absolute
tolerance of $10^{-13}$ in the 2-norm of the residual and choosing $N$
so that $|a_n|$ is below $5\cdot10^{-12}$ for $n > 3N/4$.  The number
of coefficients required strongly depends on the wavenumber $\kt$.
For example, when $0.5 \le \kt \le 4$, we find $N = 2^6$ provides
sufficient accuracy whereas for $0.002 \le \kt \le 0.01$, we use $N =
2^{12}$.

With the cosine series coefficients of $\pt(\theta)$ in hand, we
compute the unit-mean averaging integrals $\tilde{I}_j$, $j = 1,2,3$
in Eqs.~\eqref{eq:8}, \eqref{eq:whithamIntegrals} using the spectrally
accurate trapezoidal rule.  We then use sixth order finite
differencing to compute derivatives of $\tilde{I}_j$ and $\wt$ on a
grid of wavenumbers and amplitudes as explained in
Sec.~\ref{sec:large-ampl-regime}.  This numerically determines the
Whitham equations in the form \eqref{eq:whitham4}.

\subsection{Time stepping}
\label{sec:time-stepping}

\begin{figure}
\centering
\includegraphics{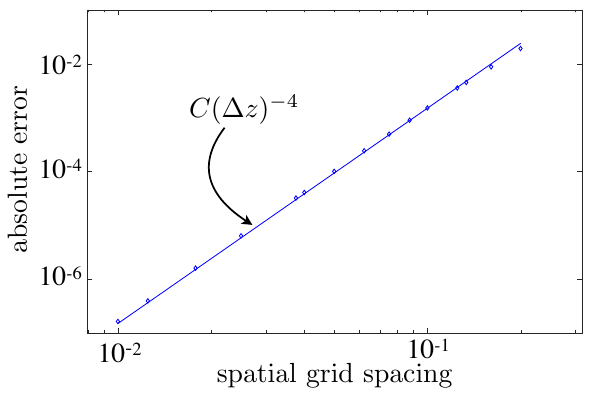}
\caption{Maximum absolute error in the direct numerical simulation of
  the conduit equation, achieving fourth order spatial accuracy as
  expected. The solution used in validation was a periodic wave with $k=3$ and $a=0.5$ generated with accuracy $10^{-8}$, and was simulated over 50 spatial periods and 5 temporal periods. The reference line is $C(\Delta z)^{-4}$.}
\label{fig:num_validation}
\end{figure}

For the direct numerical simulation of the conduit equation
\eqref{eq:conduit}, it is convenient to write it in the form of
two coupled equations:
\begin{equation}\label{eq:conduitP}
  \begin{cases}
    \Pcal=A^{-1}A_t, \\
    A\Pcal + (A^2)_z - (A^2\Pcal_z)_z = 0 .
  \end{cases}
\end{equation}
The first equation is a temporal ODE in time and the second equation
is a linear, elliptic problem $\Lcal(A)\Pcal = -(A^2)_z$ in space.  We
solve for $\Pcal$ using an equispaced fourth-order finite difference
discretization and direct inversion of the resulting banded linear
system. We implement time-dependent boundary conditions with
prescribed $A(0,t)$ and $A(L,t)$ so that the first equation in
\eqref{eq:conduitP} yields the boundary conditions for $\Pcal$.
Time-stepping is achieved with a fourth-order, explicit Runge-Kutta
method with variable timestep (\textsc{Matlab}'s \textrm{ode45}). The
solver was validated against computed periodic traveling wave
solutions. The maximum error between the numerical solution and the
periodic traveling wave solution is reported in Figure
\ref{fig:num_validation}.

\section{Nonlinear Schr\"odinger  equation derivation}
\label{sec:nonl-schr-deriv}

Here, we derive an approximation of wave modulations in the
small-amplitude, weakly nonlinear regime. Consider the ansatz
\begin{equation}
  u(z,t) = 1 + \eps u_0 + \eps^2 u_1 + \eps^3 u_2+\cdots,\, \eps\to 0,
\end{equation}
where $u_i = u_i(\theta,Z,T)$ for $i=0,1,...$, $\theta=\kt
z-\wt_0(\kt)t$, $Z=\eps z$, and $T=\eps t$, where $\wt_0(\kt)$ is the
linear dispersion relation \eqref{eq:lineardispersion} for unit mean.  Then, at
$\Order{\eps}$, we obtain a linear, homogeneous equation for $u_0$
\begin{equation}
\mathcal{L}u_0:=-\om u_{0,\theta}+2ku_{0,\theta}+k^2\om u_{0,3\theta} = 0  ,
\end{equation}
with solution $u_0 = \psi(Z,T)e^{i\theta} + c.c.$ where $c.c.$ denotes
the complex conjugate of the previous terms.  At $\Order{\eps^2}$,
$\mathcal{L}u_1 = F_1$ where 
$$F_1 = e^{2i\theta}\left[-2i \kt \psi^2\right] +
e^{i\theta}[-\psi_T-\kt^2 \psi_T-2\psi_Z+2\kt\wt_0 \psi_Z] + c.c.$$
Solvability therefore implies $ -(1+\kt^2) \left[\psi_T+\wt_0'(k)
  \psi_Z \right] \sim \eps g_1 + \cdots $, where we have introduced
the higher order correction $g_1$.  Solving for $u_1$, we include
second harmonic and mean terms $ u_1 = \psi^2(Z,T)e^{2i\theta}/(3\kt\wt)
+ c.c + M(Z,T)$ with $M$ to be determined at the next order.
Solvability with respect to constants at $\Order{\eps^3}$ yields
$$ M = \frac{(3\kt-1)(1+\kt^2)}{\kt^2(\kt^2+3)}\abs{\psi}^2. $$
Solvability with respect to the first harmonic yields $g_1$, which,
upon entering the moving reference frame and scaling to long time
$$ \xi = Z - \wt_0' T, \tau = \eps T,$$
yields the Nonlinear Sch\"odinger equation in the form
\begin{equation}
  \label{eq:conduitNLS}
  i\psi_\tau + \frac{\wt''(\kt)}{2}\psi_{\xi\xi} +
  \frac{3+5\kt^2+8\kt^4}{3\kt(\kt^2+1)(\kt^2+3)}\abs{\psi}^2\psi=0 .
\end{equation}
Appropriate scaling of $\xi$ and $\psi$ yields Eq.~\eqref{eq:NLS}.

\section{Derivation of the Whitham equations}
\label{sec:whith-equat-deriv}

For completeness, we supply a synopsis of the multiple scales
asymptotic derivation of the Whitham moduation equations.  For the
formal derivation here, we introduce slow space and time scales $Z =
\eps z$, $T = \eps t$ and consider the ansatz
\begin{equation}
  \label{eq:1}
  A(z,t) = u_0(\theta,Z,T) + \eps u_1(\theta,Z,T) + \eps^2
  u_2(\theta,Z,T) +\cdots, \quad 0 < \eps \ll 1 ,
\end{equation}
where $\theta_z = k$ and $\theta_t=-\om$. Continuity of mixed partials
$\theta_{zt} = \theta_{tz}$ implies the conservation of waves
\begin{equation}
  \label{eq:6}
    k_T + \om_Z = 0 ,
\end{equation}
one of the Whitham equations.  We insert the ansatz \eqref{eq:1} into
the conduit equation \eqref{eq:conduit} and equate like orders in
$\eps$. The $\Order{1}$ equation is
\begin{equation}
  \label{eq:MSo1}
  -\om u_{0,\theta}+2ku_0u_{0,\theta}-k^2\om
  u_{0,\theta}u_{0,\theta\theta}+k^2\om u_0u_{0,3\theta} = 0 .
\end{equation}
Equation \eqref{eq:MSo1} is solved with a family of periodic traveling
waves parameterized by $(k,a,\phib)$ (see
Sec.~\ref{sec:peri-trav-wave}) where the parameters are assumed to
depend on the slow variables $(Z,T)$.

At the next order, $\Order{\eps}$, we obtain the linear problem $\Lcal
u_1 = f$ where
\begin{align*}
  \Lcal u_1 &= -\om u_{1,\theta} + \left ( - k^2\om
    u_{0,\theta}u_{1,\theta} +2k u_0 u_1 \right)_\theta +
    k^2\om(u_{1,\theta\theta\theta}u_0 + u_{0,\theta\theta\theta} u_1 ) , \\
    f &= -u_{0,T}-k^2u_{0,\theta\theta}u_{0,T}
    +k^2u_0u_{0,\theta\theta T} - 2 u_0u_{0,Z} + 2k\om
    u_{0,\theta}u_{0,\theta Z} - 2k\om u_0u_{0,\theta\theta Z} .
\end{align*}
There are two solvability conditions in the form $ \IP{w}{f} \equiv
\int_0^{2\pi} w(\theta) f(\theta) d\theta = 0$, where $w \in
\operatorname{Ker} L^* = \operatorname{span}\{1,u_0^{-2}\}$ for the
the adjoint operator
\begin{align*}
  \Lcal^* w = \om w_\theta + k^2\om[-(u_{0,\theta}
  w)_{\theta\theta}+(u_{0,\theta\theta} w)_\theta + u_{0,3\theta} w
  -(u_0 w)_{3\theta}] +2k [u_{0,\theta} w - (u_0 w)_\theta],
\end{align*}
with $2\pi$-periodic boundary conditions.  Note that there is a third,
linearly independent function that is annihilated by $\Lcal^*$, but it
is not $2\pi$-periodic. Applying the two solvability conditions
$\IP{1}{f} = \IP{u_0^{-2}}{f} = 0$, and adding Eq.~\eqref{eq:6}, we
arrive at the Whitham equations
\begin{align}\label{eq:appendixwhitham1}
  \left(\overline{u_0}\right)_T +  \left(\overline{u_0^2}
    -2k\om\overline{u_{0,\theta}^2}\right)_Z &=0, \\ 
  \left(\overline{\frac{1}{u_0}} +
    k^2\overline{\frac{u_{0,\theta}^2}{u_0^2}} \right)_T
  -2\left(\overline{\ln u_0}\right)_Z &=0,  \\
  k_T + \om_Z &= 0
\end{align}
where $\overline{g} = \IP{1}{g}$.  Setting $\eps = 1$, i.e.,
considering the Whitham equations as the long time $t \gg 1$
asymptotic, we obtain Eqs.~\eqref{eq:whitham1}.

Averaging of the conservation laws \eqref{eq:ConsLaws} is achieved by
inserting the ansatz $A(z,t) = \phi(\theta)$ and averaging the
densities and fluxes over a period:
\begin{equation*}
  \begin{split}
    \overline{\phi}_t + \left ( \overline{\phi^2+\om
        k\phi^2 \left (\phi^{-1}\phi_\theta \right)_\theta} \right )_z&= 0, \\ 
    \left(\overline{\frac{1}{\phi} +
        k^2\frac{\phi_\theta^2}{\phi^2}}\right)_t +
    \left( \overline{-\om k\frac{\phi_{\theta\theta}}{\phi}+ \om
        k\frac{\phi_\theta\phi_\theta } {\phi^2}-2\ln{\phi}}\right)_z
    &=0 .
  \end{split}
\end{equation*}
Integration by parts and the addition of conservation of waves
\eqref{eq:6} yields the same set of Whitham equations
\eqref{eq:whitham1}.

\bibliographystyle{vancouver}


\end{document}